\documentclass[english]{article}
\usepackage[LGR,T1]{fontenc}
\usepackage[latin9]{inputenc}
\usepackage{amssymb}

\makeatletter

\DeclareRobustCommand{\greektext}{%
  \fontencoding{LGR}\selectfont\def\encodingdefault{LGR}}
\DeclareRobustCommand{\textgreek}[1]{\leavevmode{\greektext #1}}

\newcommand{\lyxaddress}[1]{
	\par {\raggedright #1
	\vspace{1.4em}
	\noindent\par}
}

\makeatother

\usepackage{babel}
\begin{document}
\title{\textbf{Quantum corrections in general relativity explored through
a GUP-inspired maximal acceleration analysis}}
\author{\textbf{$^{1,2}$Christian Corda $^{3}$Carlo Cafaro and $^{4}$Newshaw
Bahreyni}}
\maketitle

\lyxaddress{\textbf{$^{1}$SUNY Polytechnic Institute, 13502 Utica, New York,
USA }}

\lyxaddress{\textbf{$^{2}$Department of Physics and Electronics, Christ University,
Bengaluru, 560029, Karnataka, India}}

\lyxaddress{\textbf{$^{3}$University at Albany-SUNY, 12222 Albany, New York,
USA }}

\lyxaddress{\textbf{$^{4}$Pomona College, Claremont, CA 91711, USA}}

\lyxaddress{\textbf{E-mails: }\textbf{\emph{cordac.galilei@gmail.com, ccafaro@albany.edu,
Newshaw.Bahreyni@pomona.edu}}}
\begin{abstract}
A maximun acceleration analysis by Pati dating back to 1992 is here
improved by replacing the traditional Heisenberg Uncertainty Principle
(HUP) with the Generalized Uncertainty Principle (GUP), which predicts
the existence of a minimum length in Nature. This new approach allows
one to find a numerical value for the maximum acceleration existing
in Nature for a physical particle that turns out to be $a_{max}\simeq4\frac{c^{2}}{l_{P}},$
that is, a function of two fundamental physical quantities such as
the speed of light $c$ and the Planck length $l_{p}$. An application
of this result to black hole (BH) physics allows one to estimate a
new quantum limit to general relativity. It is indeed shown that,
for every real Schwarzschild BH, the maximum gravitational acceleration
occurs, without becoming infinite, when the Schwarzschild radial coordinate
reaches the gravitational radius. This means that quantum corrections
to general relativity become necessary not at the Planck scale, as
the majority of researchers in the field think, but at the Schwarzschild
scale, in agreement with recent interesting results in the literature.
In other words, the quantum nature of physics, which in this case
manifests itself through the GUP, appears to prohibit the existence
of real singularities, in this current case forbiddiing the gravitational
acceleration of a Schwarzschild BH from becoming infinite.
\end{abstract}
It is known that, in the context of a geometric reformulation of quantum
mechanics, the Heisenberg position-momentum uncertainty relation implies
the existence of a maximum acceleration for a physical particle. A
first proposal in this sense was made by Caianello and collaborators
in the first half of the 1980s \cite{key-1,key-2,key-3}. From this
first approach it emerged that, in addition to speed limits, even
higher-order rates of changes can be equally relevant in quantum physics
\cite{key-28}. In particular, Caianello obtained by means of two
different methods, the first dating back to 1981 \cite{key-1} and
the second dating back to 1984 \cite{key-3}, that the acceleration
of a particle must be limited above by the relation 

\begin{equation}
a\leq\frac{2m_{0}c^{3}}{\hslash},\label{eq:=000020Caianello}
\end{equation}
where $m_{0}$ is the mass of the particle under consideration, c
is the speed of light in vacuum and $\hslash$ is the reduced Planck
constant. This result by Caianello was criticized for three different
reasons: 
\begin{enumerate}
\item The result in Eq. (\ref{eq:=000020Caianello}) it is not a consequence
of the time-energy uncertainty principle as it was claimed by Caianello.
Instead \cite{key-4}, it arises from Landau's theory of fluctuations
\cite{key-5,key-6} and can be obtained via a standard quantum-mechanical
approach by using the \emph{ansatz} between the quantum commutator
and the classical Poisson bracket \cite{key-4}.
\item Caianello's assumption that it would be $\triangle E\leq E$ is not
always necessarily correct because, given a fixed average energy $E$,
a quantum state could, in principle, have an arbitrarily large value
of $\triangle E$ according to the uncertainty relationship between
the conjugated variables time and energy, i.e. $\triangle E\geq\frac{\hslash}{2\triangle t}$
\cite{key-7}.
\item The rest mass $m_{0}$ in Eq. (\ref{eq:=000020Caianello}) should,
in general, be replaced by the relativistic mass, $m=\gamma(v)m_{0}$
. As is well known, in Lorentz transformations the factor \textgreek{\textgamma}
depends on the speed. Furthermore \cite{key-8}, even replacing $m_{0}c^{2}$
with $mc^{2}$, one notices that for $v\rightarrow c$, $E$ is not
limited and, consequently, the acceleration of the particle will no
longer have an upper limit. 
\end{enumerate}
In 1992, Pati \cite{key-9} was clever at suggesting how to overcome
the criticisms against Caianello's approach with a derivation that
will now be reviewed. This derivation is based on the inequality in
the previously cited Landau's theory of fluctuations \cite{key-5,key-6}
\begin{equation}
\triangle E\triangle g\geq\frac{\hslash}{2}\left|\frac{dg}{dt}\right|.\label{eq:=000020inequality}
\end{equation}
Assuming that $g(t)$ in Eq. (\ref{eq:=000020inequality}) is the
position coordinate of the particle $x(t)$ and, in turn, $\left|\frac{dg}{dt}\right|\equiv v(t),$
that is, the istantaneous velocity of the particle, then one gets
\begin{equation}
\begin{array}{c}
\triangle E\triangle x\geq\frac{\hslash}{2}v\\
\\\triangle E\triangle v\geq\frac{\hslash}{2}a,
\end{array}\label{eq:=000020inequality=000020energia=000020velocit=0000E0}
\end{equation}
where $\left|\frac{dv}{dt}\right|\equiv a(t)\geq0$ is the instantaneous
acceleration of the particle. One notices that all quantities in the
second inequality in Eq. (\ref{eq:=000020inequality=000020energia=000020velocit=0000E0})
are positive, so a bit of algebra allows us to obtain 
\begin{equation}
a\leq\left(\frac{2}{\hslash}\right)^{2}\triangle E^{2}\triangle x\frac{\triangle v}{v}.\label{eq:=000020inequality=000020accelerazione}
\end{equation}
Since it is $\Delta E=\left|\frac{dE}{dp}\right|\cdot\Delta p=v\cdot\Delta p$,
one arrives at 
\begin{equation}
a\leq\left(\frac{2}{\hslash}\right)^{2}\triangle p^{2}\triangle xv\triangle v.\label{eq:=000020inequality=000020accelerazione=0000202}
\end{equation}
One notices that $v\triangle v$ is obviously bound by the relativistic
limit 
\begin{equation}
v\triangle v\leq c^{2},\label{eq:=000020v=000020delta=000020v}
\end{equation}
where $c$ is the light speed in vacuum. By considering the standard
Heisenberg uncertainty principle (HUP)
\begin{equation}
\triangle x\triangle p\geq\frac{\hslash}{2},\label{eq:=000020HUP}
\end{equation}
 one finally gets 
\begin{equation}
a\leq\frac{c^{2}}{\triangle x}\Longrightarrow a_{max}=\frac{c^{2}}{\triangle x}.\label{eq:=000020accelerazione=000020massima}
\end{equation}
It should be remembered again that the result in Eq. (\ref{eq:=000020accelerazione=000020massima})
was obtained for the first time by Pati in 1992 \cite{key-9} with
the derivation that we reviewed above. 

Now, the GUP will be used to improve this result. The GUP was originally
introduced regarding quantum gravity topics and has had various implications
in various branches of physics in recent years, see {[}10--21{]}
and the references therein. In its more general form the GUP can be
written down as \cite{key-11}
\begin{equation}
\triangle x\triangle p\geq\frac{\hslash}{2}\left[1+\beta\left(\triangle p\right)^{2}+\gamma\right],\label{eq:=000020GUP}
\end{equation}
where $\beta\equiv\beta_{0}\left(\frac{l_{P}}{\hslash}\right)^{2}$
\cite{key-19}, $l_{P}\equiv\sqrt{\frac{G\hslash}{c^{3}}}\simeq1.6\times10^{-35}\:m$
is the Planck length, $\beta_{0}$ is an adimensional constant of
the order of unit \cite{key-10} and $\gamma\equiv\beta<p>^{2}$ \cite{key-11}.
Differently from the standard HUP in Eq. (\ref{eq:=000020HUP}), the
GUP in Eq. (\ref{eq:=000020GUP}) admits the existence of a minimum
length in Nature. While in the standard HUP in Eq. (\ref{eq:=000020HUP})
$\triangle x$ can be made arbitrarily small by letting $\triangle p$
increases in correspondence, this is no longer true for the GUP in
Eq. (\ref{eq:=000020GUP}). In fact, if $\triangle p$ increases for
decreasing $\triangle x$, the new term $\beta\left(\triangle p\right)^{2}$
in the right hand side of Eq. (\ref{eq:=000020GUP}) will grow faster
than the left hand side preventing $\triangle x$ from becoming arbitrarily
small. 

By considering the state of minimum position-momentum uncertainty
relation, one rewrites Eq. (\ref{eq:=000020GUP}) as 
\begin{equation}
\triangle x=\frac{\hslash}{2}\left[\frac{1+\beta\left(\triangle p\right)^{2}+\gamma}{\triangle p}\right].\label{Dx=000020made=000020explicit}
\end{equation}
One sees that, in order to get a minimum $\triangle x,$ the first
derivative with respect to $\triangle p$ of the RHS of Eq. (\ref{Dx=000020made=000020explicit})
must be zero, that is 
\begin{equation}
\frac{\beta\left(\triangle p\right)^{2}-1-\gamma}{\left(\triangle p\right)^{2}}=0\Longrightarrow\vartriangle p=\sqrt{\frac{1+\gamma}{\beta}}.\label{eq:=000020minimum}
\end{equation}
The absolutely smallest uncertainty in position occurs when the expectation-value
of $p$, that is $<p>,$ is zero \cite{key-11}. This means 
\begin{equation}
\gamma=0\quad and\quad\vartriangle p=\sqrt{\frac{1}{\beta}}\label{eq:=000020smallest=000020uncertainty}
\end{equation}
 in Eq. (\ref{eq:=000020minimum}). By inserting the values in Eq.
(\ref{eq:=000020smallest=000020uncertainty}) into Eq. (\ref{Dx=000020made=000020explicit}),
one finds that the minimum $\triangle x$ is 
\begin{equation}
\triangle x_{m}\simeq l_{P}=\sqrt{\frac{G\hslash}{c^{3}}}\simeq1.6\times10^{-35}\:m.\label{eq:=000020Dx=000020minimo}
\end{equation}

At low energies (energies much lower than the Planck energy $E_{P}=\sqrt{\frac{\hslash c^{5}}{G}}\simeq1.2\times10^{19}GeV$),
the terms $\beta\left(\triangle p\right)^{2}$ and $\gamma$ become
irrelevant and Eq (\ref{eq:=000020GUP}) reduces to Eq. (\ref{eq:=000020HUP}).
If one now replaces the HUP in Eq. (\ref{eq:=000020HUP}) with the
GUP in Eq. (\ref{eq:=000020GUP}), one gets the maximum acceleration
as
\begin{equation}
a\leq\frac{c^{2}}{\triangle x}\left[1+\beta\left(\triangle p\right)^{2}+\gamma\right]^{2}\Longrightarrow a_{max}=\frac{c^{2}}{\triangle x}\left[1+\beta\left(\triangle p\right)^{2}+\gamma\right]^{2}.\label{eq:=000020our=000020proposal}
\end{equation}
By inserting the result in Eq. (\ref{eq:=000020minimum}), with $\gamma=0$
from Eq. (\ref{eq:=000020smallest=000020uncertainty}), and $\vartriangle x\simeq l_{P}$
in Eq. (\ref{eq:=000020our=000020proposal}), one finally obtains
\begin{equation}
a_{max}\simeq4\frac{c^{2}}{l_{P}}.\label{eq:=000020maximum=000020acceleration}
\end{equation}
By insering the standard values $c=2.998\times10^{8}\,\frac{m}{s}$
and $l_{P}\simeq1.6\times10^{-35}\:m$ in Eq. (\ref{eq:=000020maximum=000020acceleration}),
one finds the numerical value of the maximum acceleration existing
in Nature as 
\begin{equation}
a_{max}\simeq4\frac{c^{2}}{l_{P}}\simeq2.23\times10^{52}\,\frac{m}{s^{2}}.\label{eq:=000020maximum=000020value=000020acceleration}
\end{equation}
This is an enormous value, but not infinite and dependent exclusively
on the two fundamental physical quantities, speed of light and Planck
length. Clearly, accelerations so high that they tend to the very
large limiting value in Eq. (\ref{eq:=000020maximum=000020value=000020acceleration})
can only occur in extreme physical phenomena, in which very high energies
come into play. One can derive the result in Eq. (\ref{eq:=000020maximum=000020value=000020acceleration})
also in another way. Taking into account Eq. (\ref{eq:=000020smallest=000020uncertainty}),
i.e. $\gamma=0$, Eq. (\ref{Dx=000020made=000020explicit}) can be
rewritten as 
\begin{equation}
\triangle x=\frac{\hslash}{2}\frac{1}{\triangle p}+\frac{\hslash}{2}\beta\left(\triangle p\right).\label{Dx=000020made=000020explicit=0000202}
\end{equation}
Following \cite{key-20,key-21}, one introduces the Planck energy
$E_{P}\equiv\sqrt{\frac{\hslash c^{5}}{G}}$ and considers the momentum
uncertainty of the particle $\triangle p$ as being approximately
equal to the momentum $\frac{E_{P}}{c}$ of the photon with whom the
particle interacts gravitationally during a position measurement process.
Therefore, using Eq. (\ref{Dx=000020made=000020explicit=0000202}),
one gets 
\begin{equation}
\begin{array}{c}
\triangle x_{m}\simeq\frac{\hslash}{2}\frac{c}{E_{P}}+\frac{\hslash}{2}\left(\frac{l_{P}}{\hslash}\right)^{2}\frac{E_{P}}{c}\\
\\\simeq\frac{1}{2}\sqrt{\frac{G\hslash}{c^{3}}}+\frac{1}{2}\sqrt{\frac{G\hslash}{c^{3}}}\\
\\\simeq\frac{1}{2}l_{P}+\frac{1}{2}l_{P}=l_{P},
\end{array}\label{Dx=000020made=000020explicit=0000203}
\end{equation}
which is the same result as that in Eq. (\ref{eq:=000020Dx=000020minimo}).
Hence, one gets $a_{max}$ as 
\begin{equation}
\begin{array}{c}
a_{max}=\frac{c^{2}}{\triangle x_{m}}\left[1+\beta\left(\triangle p\right)^{2}\right]_{\triangle p=\frac{E_{P}}{c}}^{2}\\
\\\simeq\frac{c^{2}}{l_{P}}\left[1+\left(\frac{l_{P}}{\hslash}\right)^{2}\left(\frac{E_{P}}{c}\right)^{2}\right]^{2}\\
\\\simeq\frac{c^{2}}{l_{P}}\left[1+1\right]^{2}=4\frac{c^{2}}{l_{P}},
\end{array}\label{eq:=000020a=000020max}
\end{equation}
which is the same result as that in Eq. (\ref{eq:=000020maximum=000020value=000020acceleration}). 

Now, it is possible to show an application of this remarkable result
to BH physics. It will be shown that, for every Schwarzschild BH,
the maximum gravitational acceleration in Eqs. (\ref{eq:=000020maximum=000020acceleration})
and (\ref{eq:=000020a=000020max}) occurs, without becoming infinite,
when the Schwarzschild radial coordinate reaches the gravitational
radius. This means that quantum corrections to general relativity
become necessary not at the Planck scale, as the majority of researchers
in the field think, but at the Schwarzschild scale. Before moving
on to the actual analysis, it is worth discussing the reason why it
makes sense considering the comparison between the two accelerations
$a_{max}$ and the acceleration in general relativity, which will
be called $a_{GR}$. The maximal acceleration $a_{max}$ of the quantum
particle subject to a motion governed by quantum-mechanical rules
has been estimated using the GUP. The acceleration $a\equiv\left|<\hat{a}>\right|=\left|\frac{dv}{dt}\right|$
of the quantum particle represents, keeping in mind the fact that
the Ehrenfest Theorem 
\begin{equation}
m\frac{d^{2}<\hat{x}>}{dt^{2}}=\frac{d<\hat{p}>}{dt}=-\left\langle \overrightarrow{\nabla}V(x)\right\rangle \label{eq:=000020Ehrenfest=000020Theorem}
\end{equation}
is an expectation value version of the classical Newton's equations
of motion, the acceleration of the center of a wave-packet that moves
like a classical particle subjected to the potential $V(x)$ and Hamiltonian
\begin{equation}
H\equiv\frac{p^{2}}{2m}+V(x)\label{eq:=000020Hamiltonian}
\end{equation}
(i.e., an Hamiltonian of the same form as in classical mechanics).
The GUP, in turn, becomes important at energy scales approaching the
Planck energy with $E_{P}\approx10^{19}GeV$ (or, alternatively, $M_{P}\approx10^{-8}Kg$).
The gravitational effects of high energies (i.e., $\gtrsim10^{19}GeV$)
attempting to resolve small distances will ultimately cause significant
disturbance to the space-time structure. At energies much below $E_{P},$
the GUP becomes the standard HUP, where gravitational effects are
negligible compared to quantum effects. Interestingly, despite the
fact that the standard Newtonian gravitational theory is sufficient
to arrive at the essential meaning of the GUP, a more formal derivation
of the GUP relies upon a fully general relativistic description of
the interaction between the photon and the test particle being observed
(i.e., whose position is being measured) \cite{key-13}. Therefore,
to model the departure from a flat spacetime to a curved spacetime,
one regards the quantum particle as a test particle in a Schwarzschild
space-time. Then, considering the general relativistic force on such
a test particle, one introduces the acceleration $a_{GR}$. Therefore,
for internal consistency, one imposes $\left|a_{GR}\right|\leq a_{max}$.

One starts by remembering that the radial acceleration of a test mass
at a Schwarzschild radial coordinate $r$ from a Schwarzschild BH
is \cite{key-22}
\begin{equation}
a_{GR}=-\frac{GM}{r^{2}\sqrt{1-\frac{2GM}{c^{2}r}}},\label{eq:=000020Schwarzschild=000020radial=000020acceleration}
\end{equation}
where $M$ and 
\begin{equation}
r_{g}\equiv\frac{2GM}{c^{2}},\label{eq:=000020gravitational=000020radius}
\end{equation}
are the BH mass and the BH gravitational radius, respectively \cite{key-23}.
One observes that: 

i) In case of weak gravitational field, to first order in the gravitational
potential $-\frac{GM}{r}$, the acceleration in Eq. (\ref{eq:=000020Schwarzschild=000020radial=000020acceleration})
reduces to the Newtonian one \cite{key-22}
\begin{equation}
a_{Newtonian}\simeq-\frac{GM}{r^{2}}\label{eq:=000020Newtonian=000020acceleration}
\end{equation}

ii) The Einstein Equivalence Principle implies that in a local inertial
reference frame, i.e. a reference frame in which the test mass is
in free fall, the gravitational field vanishes, and therefore the
corresponding acceleration also vanishes \cite{key-22,key-23}. In
this case, in order to talk about acceleration of the test mass, one
must imagine that there is something that holds the test mass at rest,
opposing its free fall. One can then hypothesize that the test mass
is on the floor of a spaceship blocked by an imaginary winch that
opposes the gravitational field. In this way, the observer in solidarity
with the test mass experiences a sense of possessing weight. 

iii) In the limit in which $r\rightarrow r_{g}$ the acceleration
in Eq. (\ref{eq:=000020Schwarzschild=000020radial=000020acceleration})
diverges. 

Now, the limit in Eqs. (\ref{eq:=000020maximum=000020value=000020acceleration})
and (\ref{eq:=000020a=000020max}) implies that 
\begin{equation}
a_{max}\simeq4\frac{c^{2}}{l_{P}}\gtrsim\frac{GM}{r^{2}\sqrt{1-\frac{2GM}{c^{2}r}}}.\label{eq:=000020limite}
\end{equation}
Remembering that the definition of gravitational (Schwarzschild) radius
in Eq. (\ref{eq:=000020gravitational=000020radius}), Eq. (\ref{eq:=000020limite})
becomes 
\begin{equation}
\frac{4}{l_{P}}\gtrsim\frac{r_{g}}{2r^{2}\sqrt{1-\frac{r_{g}}{r}}},\label{eq:=000020limite=0000202}
\end{equation}
and, for $r>r_{g}$, one finds
\begin{equation}
r^{2}\sqrt{1-\frac{r_{g}}{r}}\gtrsim\frac{r_{g}l_{P}}{8}.\label{eq:=000020limite=0000203}
\end{equation}
Therefore, since for $r>r_{g}$, both sides of Eq. (\ref{eq:=000020limite=0000203})
are real and positive, by squaring one gets 
\begin{equation}
r^{4}-r_{g}r^{3}-\left(\frac{r_{g}l_{P}}{8}\right)^{2}\gtrsim0.\label{eq:=000020limite=0000204}
\end{equation}
By taking $r=\alpha r_{g,}$where $\alpha>1$ is a real parameter,
one considers the function 
\begin{equation}
y\left(r=\alpha r_{g}\right)\equiv r_{g}^{2}\left[r_{g}^{2}\alpha^{3}\left(\alpha-1\right)-\frac{l_{P}^{2}}{64}\right].\label{eq:=000020y}
\end{equation}
One gets
\begin{equation}
\begin{array}{ccc}
y\left(r=\alpha r_{g}\right)\gtrsim0 & for & r_{g}\gtrsim\frac{l_{P}}{8\sqrt{\alpha^{3}\left(\alpha-1\right)}},\end{array}\label{eq:=000020quasi-positivity=000020of=000020y}
\end{equation}
and, in particular,
\begin{equation}
\begin{array}{ccc}
y\left(r=\alpha r_{g}\right)\simeq0 & for & r_{g}\simeq\frac{l_{P}}{8\sqrt{\alpha^{3}\left(\alpha-1\right)}},\end{array}\label{eq:=000020zero=000020of=000020y}
\end{equation}
where $r=\alpha r_{g}$ expresses the Schwarzschild radial coordinate
$r$ in terms of the gravitational radius $r_{g}$. When the function
$y\left(r=\alpha r_{g}\right)$ vanishes in Eq. (\ref{eq:=000020y}),
the corresponding Schwarzschild radial coordinate $r=\alpha r_{g}$
gives the point where the BH gravitational acceleration reaches its
maximum given by Eqs. (\ref{eq:=000020maximum=000020acceleration})
and (\ref{eq:=000020a=000020max}). Then, for $\alpha=2$ one obtains
\begin{equation}
\begin{array}{ccc}
y\left(r=2r_{g}\right)\gtrsim0 & for & r_{g}\gtrsim\frac{l_{P}}{16\sqrt{2}}\simeq0.044l_{P}.\end{array}\label{eq:=000020condition}
\end{equation}
The condition $>$ in Eq. (\ref{eq:=000020condition}) is satisfied
for \emph{all} Schwarzschild BHs since it is well known that the smallest
existing BH has gravitational radius of the order of the Planck length.
The meaning of Eq. (\ref{eq:=000020condition}) is then that at the
distance $r=2r_{g}$ the maximum acceleration is reached only for
BHs with gravitational radius of the order of $0.044l_{P}$, which
do not exist in Nature, while for BHs with a greater gravitational
radius the maximum acceleration at the distance $r=2r_{g}$ is never
reached. To reach $a_{max}$, one needs to get closer to the gravitational
radius. Inserting $\alpha=1.5$ into Eq. (\ref{eq:=000020y}), one
finds that, when the Schwarzschild radial coordinate is $r=1.5r_{g},$
one obtains 
\begin{equation}
\begin{array}{ccc}
y\left(r=1.5r_{g}\right)\gtrsim0 & for & r_{g}\gtrsim0.096l_{P}.\end{array}\label{eq:=000020condition=0000201.5}
\end{equation}
Thus, again, the meaning of Eq. (\ref{eq:=000020condition=0000201.5})
is that, at the Schwarzschild radial coordinate $r=1.5r_{g},$ the
maximum acceleration is reached only for BHs with gravitational radius
of the order of $0.096l_{P}$, which do not exist in Nature, also
because the GUP tells one that the Planck length $l_{P}$ is the minimum
length existing in Nature, while for BHs with a greater gravitational
radius the maximum acceleration at the distance $r=1.5r_{g}$ is never
reached. To reach $a_{max}$, one needs to get closer to the Schwarzschild
radius. Inserting $\alpha=1.01$ into Eq. (\ref{eq:=000020y}), one
finds that, when the Schwarzschild radial coordinate is $r=1.01r_{g},$
it is 
\begin{equation}
\begin{array}{ccc}
y\left(r=1.01r_{g}\right)\gtrsim0 & for & r_{g}\gtrsim1.232l_{P}.\end{array}\label{eq:=000020condition=0000201.01}
\end{equation}
BHs of this size (i.e. of the order of the Planck length $l_{P}$),
could exist in Nature. Thus, it has been found that the maximum gravitational
acceleration from these BHs, that is, the one in Eqs. (\ref{eq:=000020maximum=000020acceleration})
and (\ref{eq:=000020a=000020max}), is achieved at the Schwarzschild
radial coordinate $r=1.01r_{g}$. For BHs with a greater gravitational
radius the maximum acceleration at the distance $r=1.01r_{g}$ is
never reached. Therefore, one needs to get even closer to the Schwarzschild
radius. Inserting $\alpha=1.0001$ into Eq. (\ref{eq:=000020y}),
one sees that when the Schwarzschild radial coordinate is $r=1.0001r_{g},$
one obtains 
\begin{equation}
\begin{array}{ccc}
y\left(r=1.0001r_{g}\right)\gtrsim0 & for & r_{g}\gtrsim12.498l_{P}.\end{array}\label{eq:=000020condition=0000201.0001}
\end{equation}
 Then, it has been found that the maximum gravitational acceleration
from these BHs, that is, the one in Eqs. (\ref{eq:=000020maximum=000020acceleration})
and (\ref{eq:=000020a=000020max}), is achieved at the Schwarzschild
radial coordinate $r=1.0001r_{g}.$ For BHs with a greater gravitational
radius than the one in Eq. (\ref{eq:=000020condition=0000201.0001}),
the maximum acceleration at the distance $r=1.0001r_{g}$ is never
reached. Thus, one needs to get even closer to the gravitational radius.
One also notes that it is 
\begin{equation}
\lim_{\alpha\rightarrow1^{+}}\frac{l_{P}}{8\sqrt{\alpha^{3}\left(\alpha-1\right)}}=+\infty,\label{eq:=000020Limit}
\end{equation}
which means that the maximum gravitational acceleration in Eqs. (\ref{eq:=000020maximum=000020acceleration})
and (\ref{eq:=000020a=000020max}), is reached at the event horizon
only for BHs approximating an infinite Schwarzschild radius (and therefore
infinite mass). Furthermore, the closer $\alpha$ gets to unity in
the function $\frac{l_{P}}{8\sqrt{\alpha^{3}\left(\alpha-1\right)}}$
, the more the performance of the same function will depend on $(\alpha-1)$.
Thus, also considering the previous results, one argues that every
real BH will reach the maximum gravitational acceleration in Eqs.
(\ref{eq:=000020maximum=000020acceleration}) and (\ref{eq:=000020a=000020max})
when the Schwarzschild radial coordinate approaches the event horizon.
The larger are the BH gravitational radius and the BH mass, the closer
to the event horizon will be the Schwarzschild radial coordinate at
which the BH maximum gravitational acceleration in Eqs. (\ref{eq:=000020maximum=000020acceleration})
and (\ref{eq:=000020a=000020max}) is reached. On the other hand,
it will now be demonstrated that, in reality, the distance (understood
as the difference between two different Schwarzschild radial coordinates)
between the point of maximum acceleration and the event horizon cannot
be \textquotedbl resolved\textquotedbl{} as it is always less than
the Planck length. Setting $k\equiv\frac{r_{g}}{l_{P}}$, if one squares
the second inequality in Eq. (\ref{eq:=000020quasi-positivity=000020of=000020y}),
one gets 
\begin{equation}
\alpha^{3}\left(\alpha-1\right)\gtrsim\frac{1}{64k^{2}}.\label{eq:=000020second=000020way}
\end{equation}
The first derivative of 
\begin{equation}
F(\alpha)\equiv\alpha^{3}\left(\alpha-1\right)-\frac{1}{64k^{2}}=\alpha^{4}-\alpha^{3}-\frac{1}{64k^{2}}\label{eq:=000020F=000020of=000020alpha}
\end{equation}
is 
\begin{equation}
F'(\alpha)\equiv4\alpha^{3}-3\alpha^{2}=\alpha^{2}\left(4\alpha-3\right).\label{eq:=000020first=000020derivative}
\end{equation}
Thus, $F'(\alpha)$ is always positive for $\alpha>\frac{3}{4}$,
which means that $F(\alpha)\,$ always increases for $\alpha>\frac{3}{4}$.
On the other hand, $F(\alpha)$ is negative for $\alpha=1,$ while
it is $\lim_{\alpha\rightarrow+\infty}F(\alpha)=+\infty.$ Then, for
every $k,$ there will be a unique value of $\alpha$ for which $F(\alpha)$
vanishes in the interval $1<\alpha<+\infty.$ In order to find this
value one sees that, for $\alpha\simeq1$ it is $\alpha^{3}\left(\alpha-1\right)\simeq\alpha-1$.
Thus, Eq. (\ref{eq:=000020second=000020way}) becomes 
\begin{equation}
\alpha\gtrsim1+\frac{1}{64k^{2}}.\label{eq:=000020second=000020way=0000202}
\end{equation}
Hence, the unique approximate value of $\alpha_{0}$ for which $F(\alpha)$
vanishes in the interval $1<\alpha<+\infty$ for every $k$ is 
\begin{equation}
\alpha_{0}\simeq1+\frac{1}{64k^{2}}.\label{eq:=000020second=000020way=0000203}
\end{equation}
The value of $\alpha_{0}$ in Eq. (\ref{eq:=000020second=000020way=0000203})
is the value corresponding to the maximum gravitational acceleration
in Eqs. (\ref{eq:=000020maximum=000020acceleration}) and (\ref{eq:=000020a=000020max})
for each fixed value of the ratio between the BH gravitational radius
and the Planck lenght. From Eq. (\ref{eq:=000020second=000020way=0000203})
it is understood that the more the BH mass increases, the more $\alpha_{0}\rightarrow1,$
since $k\equiv\frac{r_{g}}{l_{P}}=\frac{2M}{M_{P}}$ and $k^{-2}\propto M^{-2}$,
with $M$ being the BH mass. This confirms our previous result that
every Schwarzschild BH will reach the maximum gravitational acceleration
in Eqs. (\ref{eq:=000020maximum=000020acceleration}) and (\ref{eq:=000020a=000020max})
when the Schwarzschild radial coordinate approaches the gravitational
radius. The larger the BH gravitational radius and the BH mass are,
the closer to the Schwarzschild radius will be the Schwarzschild radial
coordinate at which the BH maximum gravitational acceleration in Eqs.
(\ref{eq:=000020maximum=000020acceleration}) and (\ref{eq:=000020a=000020max})
is reached. For stellar BHs it is $k\sim10^{39}.$ Thus, $\alpha_{0}\simeq1+10^{-80}.$
In this case, the BH gravitational radius is of the order of tens
of kilometers and therefore the distance between the point of maximum
acceleration and the BH horizon is many orders of magnitude smaller
than the Planck length. It is concluded that, in fact, by the maximum
gravitational acceleration for stellar BHs is reached at the moment
a test mass meets the horizon, without becoming infinite. Considering
Eq. (\ref{eq:=000020second=000020way=0000203}), one observes that
it is also 
\begin{equation}
\begin{array}{ccccc}
\left(\alpha_{0}-1\right)r_{g}\simeq l_{P} & for & \frac{1}{64k^{2}}\simeq\frac{1}{k} & \Longrightarrow & k\simeq\frac{1}{64}\end{array}.\label{eq:=000020about}
\end{equation}
Obviously, BHs that have the ratio of gravitational radius to Planck
length of the order of $1/64$ do not exist in Nature. One therefore
concludes that what it has been found for stellar BHs is valid for
all Schwarzschild BHs existing in Nature. That is, because of the
GUP, one cannot \textquotedbl solve\textquotedbl{} the difference
between the Schwarzschild radial coordinate at which the maximum acceleration
is reached and the gravitational radius for all Schwarzschild BHs.
In other words, the distance between the point of maximum acceleration
and the BH horizon is always smaller than the Planck length. It should
also be remembered that, in a quantum framework, a BH horizon is not
static, but is instead subject to quantum fluctuations, whether it
is considered a real horizon \cite{key-29} or an apparent horizon
\cite{key-26}, as it has been demonstrated in quantum approaches
to gravitational collapse \cite{key-25,key-26}. Even from this other
point of view, it therefore makes no sense to talk about a distance
of a point from the BH horizon that is smaller than the Planck length.
Thus, one concludes that, in fact, the maximum gravitational acceleration
for all Schwarzschild BHs is reached at the moment the test mass meets
the horizon, without becoming infinite.

Thus, the following intriguing result has been found: For every Schwarzschild
BH, the maximum gravitational acceleration occurs, without becoming
infinite, when the Schwarzschild radial coordinate reaches the gravitational
radius. This means that quantum corrections to general relativity
become necessary not at the Planck scale, as the majority of researchers
in the field think, but at the Schwarzschild scale, in perfect agreement
with some interesting results of recent years on BH physics, see for
example \cite{key-24,key-25,key-26}. In other words, the quantum
nature of physics, which in the current case manifests itself through
the GUP, appears to prohibit the existence of real singularities,
in this case forbidding the gravitational acceleration of a Schwarzschild
BH from becoming infinite.

\subsection*{From non-relativistic to relativistic GUP}

Our analysis relies upon the non-relativistic GUP. However, we wish
to underscore that there have been recent criticisms directed towards
this form of GUP \cite{key-30}. Specifically, it is important to
note that the a fully-relativistic GUP must fulfill the following
three conditions {[}30-32{]}:

(i) Preserving Lorentz covariance;

(ii) Preserving the linearity of the dispersion relations;

(iii) Relating the uncertainty principle to the spacetime metric.

In future research efforts, it will be beneficial to investigate the
relativistic generalized uncertainty principle (RGUP), initially presented
in {[}32-34{]}, and strive to enhance the findings of this paper.
However, conversely, certain issues highlighted in \cite{key-30}
can, in theory, be addressed as follows. The non-Lorentz invariance
of the GUP does not pose an issue when we examine the minimum length
from the perspective of the accelerated body, as in special relativity
the time and length experienced by the moving observer are always
shorter than those measured by the fixed observer. This represents
an absolute minimum length.

Similarly, the requirement for the GUP to be linked to the space-time
metric appears to be unnecessary in this context, as these are local
measurements due to the infinitesimal distances involved and Einstein's
equivalence principle indicates that space-time is consistently locally
flat. Conversely, scientific progress occurs gradually, and the findings
presented in this paper represent a noteworthy initial step. However,
we concur with Ref. {[}30{]} that a more comprehensive analysis would
necessitate the application of a relativistic GUP, which falls outside
the scope of the present study. Ultimately, it is crucial to remember
that a recent advancement indicates that a new prediction regarding
maximal proper force also corroborates the forecast of maximal proper
acceleration \cite{key-30}. Gravitational force represents the maximum
proper force, which is associated with supplementary (quantum) curvatures.
A recent example can be observed in {[}35, 36{]}, within a quantum
geometric framework that may provide an alternative mathematical structure
for comprehending the emergence of quantum gravity and addressing
the singularity dilemma in general relativity.

\subsection*{Concluding remarks}

In summary, in this letter an acceleration analysis by Pati dating
back to 1992 has been generalized by replacing the traditional HUP
with the GUP which predicts the existence of a minimum length in Nature.
This new approach allows us to estimate a numerical value for the
maximum acceleration existing in Nature for a physical particle which
turns out to be $a_{max}\simeq4\frac{c^{2}}{l_{P}}\simeq2.23\times10^{52}\,\frac{m}{s^{2}},$
i.e. a very large value, but finite, and a function of two fundamental
physical quantities such as the speed of light $c$ and the Planck
length $l_{p}.$ Then, an application of this remarkable result to
BH physics has been discussed. In fact, it has been shown that, for
every Schwarzschild BH, the maximum gravitational acceleration occurs,
without becoming infinite, when the Schwarzschild radial coordinate
reaches the gravitational radius. This means that quantum corrections
to general relativity become necessary not at the Planck scale, as
the majority of researchers in the field think, but at the Schwarzschild
scale, in agreement with some interesting results of recent years
on BH physics \cite{key-24,key-25,key-26}. It is better to clarify
the meaning of this last statement. In classical general relativity,
the inequality $\left|a_{GR}\right|\leq a_{max}$ is evidently violated,
because the acceleration becomes infinite for $r=r_{g}$, but, if
one assumes that the nature of the world is quantum, one must \textquotedbl force\textquotedbl{}
the particle to maintain a finite acceleration, that is, impose that
the inequality $\left|a_{GR}\right|\leq a_{max}$ is NEVER violated.
Calculations tell one that this maximum acceleration is reached at
$r\simeq r_{g}$, that is at the Schwarzschild scale, long before
reaching the Planck scale. In this sense, therefore, quantum corrections
to general relativity seem necessary at the Schwarzschild scale. Therefore,
in the particular case discussed in this letter, the quantum nature
of physics, which in the present approach manifests itself through
the GUP, appears to prohibit the existence of real singularities,
by forbidding the gravitational acceleration of a Schwarzschild BH
from becoming infinite.

For the sake of completeness, it is worth mentioning the first work,
to the knowledge of the Authors, in which there was a discussion of
a possible minimum length in Nature. This is a pioneering work by
Mead \cite{key-27}, dating back to 60 years ago (1964, that is, long
before there was some discussion about a GUP in the literature), in
which the Author analyzes a gravitational effect in quantum mechanics
(uncertainty principle) which shows that it is impossible to measure
the position of a particle with error less than $\triangle x_{m}\gtrsim1.6\times10^{-35}m$,
which is exactly the Planck length in Eq. (\ref{eq:=000020Dx=000020minimo}). 

For the sake of thoroughness, it is important to note that although
Caianiello and his colleagues were indeed the originators of the concept
of maximum proper acceleration, it was Brandt in Ref. \cite{key-37}
who further developed this analysis by incorporating the ideas of
spacetime and the generalized uncertainty principle. Specifically,
both research groups also contemplated the idea of a generalized form
of eight-dimensional spacetime, which includes adaptations to Finsler
geometry contexts.

Ultimately, as highlighted in Reference \cite{key-30}, we stress
the importance of investigating the RGUP {[}32-34{]} in future research
articles and, ideally, endeavoring to enhance the findings presented
in this paper.

\section*{Acknowledgements}

The Authors wish to extend their appreciation to an anonymous Reviewer
for insightful advice and recommendations, particularly for emphasizing
the RGUP that will be employed in a forthcoming research effort.

\end{document}